\documentclass[%
 reprint,
 superscriptaddress,
 amsmath,amssymb,
 aps,
 prl,
floatfix,
]{revtex4-2}

\usepackage{physics}
\usepackage{graphicx}
\usepackage{xcolor}

\usepackage{pdfpages} 
\usepackage{pgffor} 

\makeatletter
\AtBeginDocument{\let\LS@rot\@undefined}
\makeatother

\def\supplementfilename{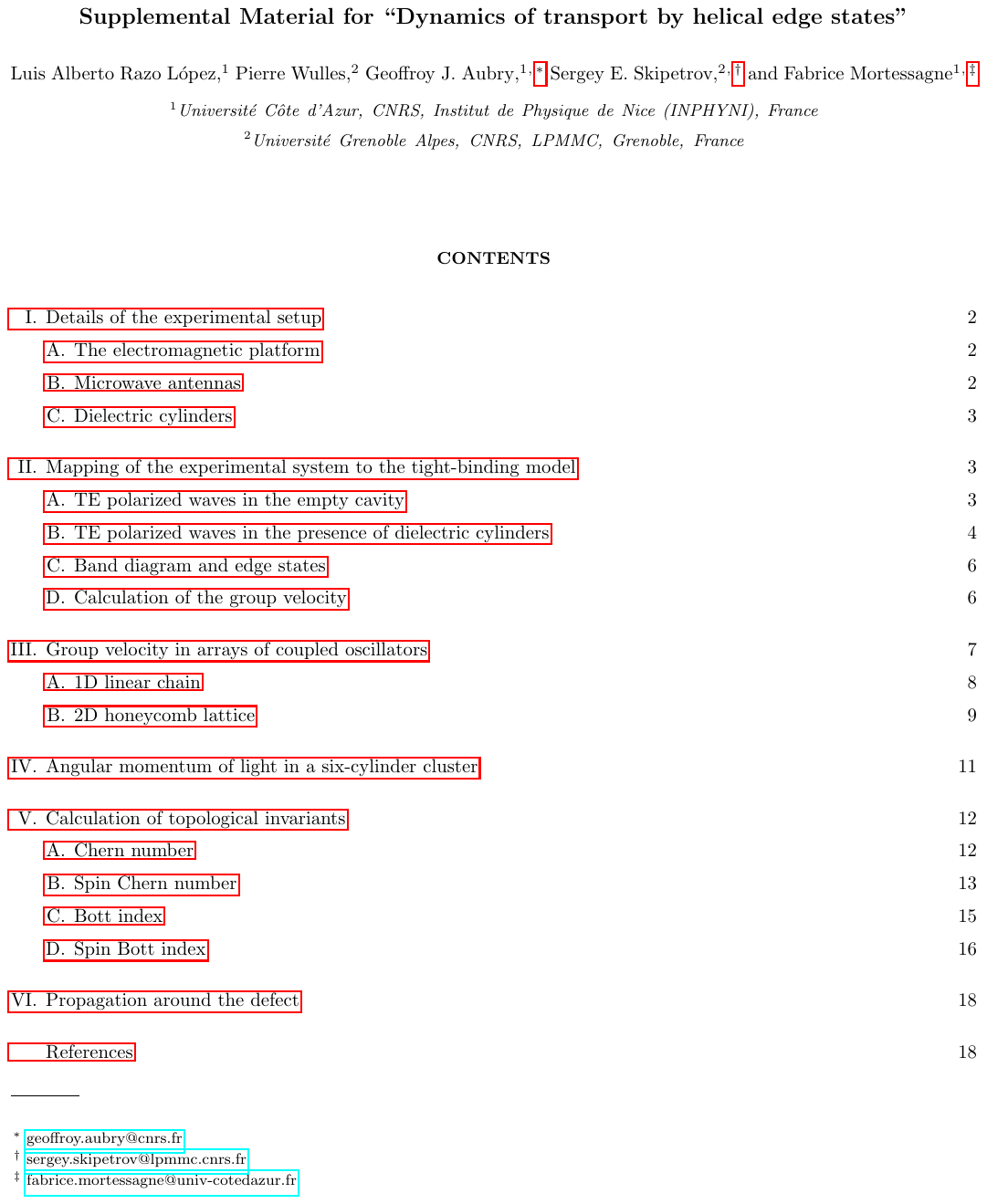} 

\pdfximage{\supplementfilename}
\def\numbersupplementpages{\the\pdflastximagepages}

\newif\ifarXiv
\arXivtrue 

\usepackage{hyperref}

\renewcommand{\vec}[1]{{\mathbf #1}}

\begin{document}
\title{Dynamics of transport by helical edge states}

\author{Luis Alberto Razo López}
\affiliation{Universit\'e Côte d'Azur, CNRS, Institut de Physique de Nice (INPHYNI), France}
\author{Pierre Wulles}
\affiliation{Universit\'e Grenoble Alpes, CNRS, LPMMC, Grenoble, France}
\author{Geoffroy J. Aubry}
\email{geoffroy.aubry@cnrs.fr}
\affiliation{Universit\'e Côte d'Azur, CNRS, Institut de Physique de Nice (INPHYNI), France}
\author{Sergey E. Skipetrov}
\email{sergey.skipetrov@lpmmc.cnrs.fr}
\affiliation{Universit\'e Grenoble Alpes, CNRS, LPMMC, Grenoble, France}
\author{Fabrice Mortessagne}
\email{fabrice.mortessagne@univ-cotedazur.fr}
\affiliation{Universit\'e Côte d'Azur, CNRS, Institut de Physique de Nice (INPHYNI), France}

\begin{abstract}
Topologically nontrivial band structure of a material may give rise to special states that are confined to the material's boundary and protected against disorder and scattering. Quantum spin Hall effect (QSHE) is a paradigmatic example of phenomenon in which such states appear in the presence of time-reversal symmetry in two dimensions. Whereas the spatial structure of these helical edge states has been largely studied, their dynamic properties are much less understood. We design a microwave experiment mimicking QSHE and explore the spatiotemporal dynamics of unidirectional transport of optical angular momentum (or pseudospin) by edge states. Pseudospin-polarized signal propagation is shown to be immune to scattering by defects introduced along the edge. 
Its velocity is 2 to 3 orders of magnitude slower than the speed of light in the free space, which may have important consequences for practical applications of topological edge states in modern optical and quantum-information technologies.
\end{abstract}

\maketitle

Quantum Hall effect (QHE) \cite{klitzing80,klitzing86} as well as its variants---quantum spin Hall \cite{kane05a,konig07} and quantum anomalous Hall \cite{yu10,chang13} effects (QSHE and QAHE)---are archetypal examples of topological phenomena in condensed matter physics. They are due to the nontrivial topological structure of electronic bands in some two-dimensional (2D) materials, leading to the appearance of robust conducting edge states protected against scattering by defects and giving rise to a quantized conductance \cite{oh13,bernevig13}. The direct relationship between edge states and quantized conductance has been challenged by recent experiments in which electric current density in QAHE has been measured with spatial resolution and found to be not restricted to sample edges \cite{rosen22,ferguson23}. Together with theoretical modeling \cite{doucot24}, such measurements unveil unexpected microscopic details of the robust conductance quantization in QAHE and demonstrate that our current understanding of topological phenomena in condensed matter physics is far from being complete. 

Whereas spatially resolved measurements represent a real experimental challenge in 2D electron gases, they are quite routinely performed in photonic or acoustic setups designed to mimic electronic systems. Indeed, classical-wave analogs of quantum Hall effects have been proposed and experimentally demonstrated rather soon after their discoveries, first in photonics \cite{haldane08,wang09,lu14,ozawa19} and then in acoustics \cite{ma19,xue22}. Spatial maps of eigenmodes or wave fields in topologically nontrivial samples have been presented in virtually every publication in this research field. Cold-atom experiments in which the role of 2D electron gas is played by an ultracold atomic gas also allow for measurements with spatial resolution \cite{jotzu14,cooper19,braun24,yao24}. In addition to facilitating position-resolved measurements, classical-wave and cold-atom systems have the advantage of being easier to control, allowing for observing the investigated physical phenomena in their purest form. In particular, interactions between particles (photons, phonons or atoms) can be made negligibly weak, disorder can be finely tuned, and finite-temperature effects can be almost fully suppressed. This makes such systems interesting to clarify details of transport phenomena in quantum Hall effects by going beyond the transverse and longitudinal conductivity measurements that are common in electron transport experiments.

\begin{figure*}[t]
    \centering
        \includegraphics{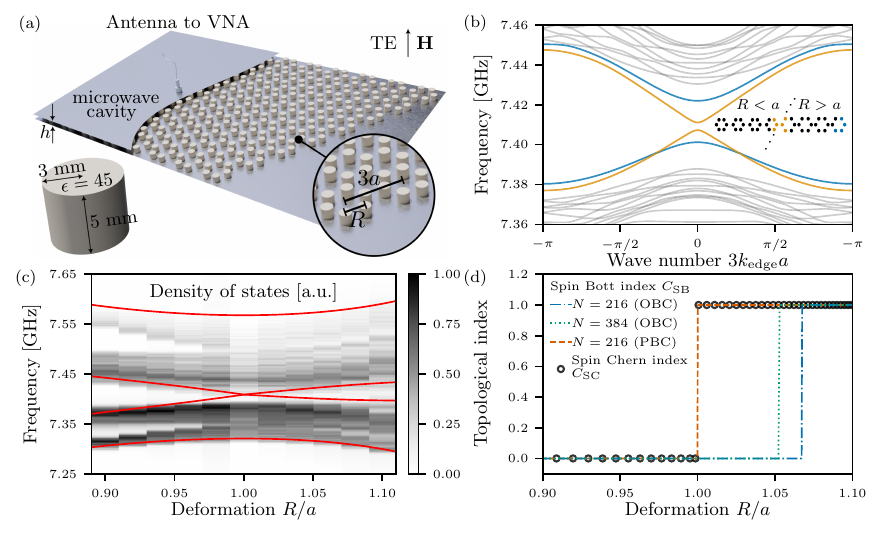}
     \caption{
     (a) Experimental setup. $N = 384$ dielectric cylinders of 3 mm in radius and 5 mm in height, dielectric constant $\epsilon = 45$, are grouped in six-cylinder hexagonal clusters of side $R$.
     In their turn, the clusters are arranged in a triangular lattice with $3a = 30$ mm lattice spacing and placed on the lower of the two reflecting aluminum plates separated by a distance of $h = 13$ mm and forming a Fabry-Perot cavity with no propagating TE modes.
    TE-polarized microwaves are emitted by the first of the two loop antennas introduced into the cavity via small holes in the plates. The second antenna can move together with the upper plate.
    (b) Band structure of an infinite ribbon (see inset for a sketch of its single unit cell repeated along dotted lines) divided in two halves with $R < a$ on the left and $R > a$ on the right, and having the same types of boundaries as our sample, in the tight-binding approximation. $\vec{k}_{\text{edge}}$ is parallel to ribbon edges. Color lines show bands corresponding to the edge states at the interface between the two halves of the sample (yellow) and at the sample boundary (blue).
    A (mini-)gap between edge states is due to the breakdown of the $C_6$ symmetry at an edge.
    Calculations done using PythTB~\cite{Coh_2024}.
    (c) Gray-scale plot of measured DOS compared to a calculation of band edges in the tight-binding model (red lines), as a function of frequency and deformation $R/a$.
    The apparent persistence of the gap in the measured DOS for $R/a = 1$ is due to finite frequency resolution of our measurements.
    (d) Spin Chern number $C_{\text{SC}}$ for the infinite lattice (circles) and spin Bott index $C_{\text{SB}}$ for lattices of two different sizes and open (OBC) or periodic (PBC) boundary conditions. Both $C_{\text{SC}}$ and $C_{\text{SB}}$ are calculated for the middle of the spectral gap in panel (c).
    }
     \label{fig1}
\end{figure*}

In contrast to the spatial structure of transport in topologically nontrivial systems, its dynamical aspects are largely unexplored to date. First measurements of propagation velocities associated with topologically protected edge states have been performed in ultracold quantum gases only very recently~\cite{braun24,yao24}. These works deal with {\it chiral} edge states that arise in systems with broken time-reversal (TR) symmetry and can be roughly thought of as analogous to those in QHE or QAHE.
Studies of dynamical aspects of transport by {\it helical} edge states in TR-symmetric QSHE-like systems are scarce.
A notable example of such a study demonstrated helical transport in a network of pairs of coupled mechanical oscillators \cite{Suesstrunk2015}. 
In the present work, we perform experiments in a 2D microwave setup mimicking QSHE, see Fig.~\ref{fig1}~(a) and Supplemental Material (SM)~\footnote{
See SM that contains information on the experimental setup, details on the mapping between the experimental system and the tight-binding model, the computation of the group velocity in arrays of coupled oscillators, the computation of the angular momentum of light in six-cylinder clusters, and the computation of the topological invariants. The SM includes Refs.~\cite{Jackson_1999, Reisner_PhD_2023,Chern1946, fukui05, prodan11prb, prodan11, Griffiths2018Aug, loring10, Loring2019Jul, bianco11, Huang2018prl, Wulles2024}.} Sec.~I.
In contrast to Ref.~\cite{Suesstrunk2015} where an intricate and purposely designed coupling between oscillators was required, our setup relies on natural electromagnetic coupling between dielectric resonators. The role of spin---the intrinsic angular momentum of electrons---is played by the orbital angular momentum $\vec{J}$ of electromagnetic waves. It originates from the microwave energy flow in hexagonal clusters (side $R$) of weakly coupled microwave resonators (dielectric cylinders) playing the role of elementary cells and arranged in a triangular lattice of spacing $3a$~\cite{wu15}.
The symmetry behind topological phenomena in our experiments is the (pseudo-)TR one: an analog of the TR operator is constructed as $T = UK$, where $U = i\sigma_z$ with $\sigma_z$ being the Pauli matrix, and $K$ is the complex conjugation \cite{wu16}.
Due to the $C_6$ symmetry of the hexagonal six-cylinder structure, its $p$ and $d$ orbitals are eigenvectors of $T$ with $T^2 = -1$, which produces Kramers degeneracy of bands at the analog of the Dirac point in the band structure of our experimental system when $R = a$ and cylinders form a honeycomb lattice.
Expanding six-cylinder clusters ($R > a$) opens a topological gap in the spectrum at the Dirac point, with edge states crossing the gap in a sample with boundaries \cite{wu16}.
Breakdown of $C_6$ symmetry near sample edges spoils the analogy between the pseudospin $\vec{J}$ and the electron spin, leading to the opening of a narrow spectral mini-gap between the bands corresponding to edge states [see Fig.~\ref{fig1}~(b)] and to the mixing of the two pseudospin channels ~\cite{wu15,wu16}.
As convincingly demonstrated in previous experiments with microwaves \cite{yves17,yang18} and light \cite{noh18,barik18,para20}, this does not compromise the topological nature of the band structure or the helicity of edge modes but limits the distance over which pseudospin-polarized propagation along sample edges can be observed.
The focus of our work resides in position- and time-resolved measurements of the pseudospin $\vec{J}$, allowing us to directly observe unidirectional propagation of microwaves and pseudospin transport along an interface between topologically distinct parts of the sample.
We demonstrate that the transport of pseudospin is not perturbed by defects introduced at the interface and measure both the group velocity of waves propagating along the interface and the velocity of pseudospin transport. 

In our experiments, microwave reflection $S_{11}(\vec{r},f)$ and transmission $S_{21}(\vec{r},f)$ via TE modes of the lattice of dielectric cylinders are measured using two loop antennas, one of which is fixed in the middle of the sample and the other one is placed at a position $\vec{r}$~\cite{Bellec_PRB_2013, reisner21, RazoLopez2024} (see Fig.~\ref{fig1}~(a) and SM~\cite{Note1}~Sec.~I). 
The explored frequency range $f = 7.1$--7.7 GHz around the first magnetic-dipole resonance of an individual cylinder is below the cutoff $c/2h$ of the first TE mode of the empty Fabry-Perot cavity hosting the sample (where $c$ is the speed of light in the free space), so that electromagnetic excitations only exist inside the cylinders and the latter are coupled through the spatial overlap of evanescent waves.
Microwave propagation in such a system can be mapped on a tight-binding model with nearest-neighbor couplings $\tau(d)$ that decrease exponentially with the distance $d$ between cylinders~\cite{bellec13}.
We define $\tau_\mathrm{in} = \tau(R)$ as the coupling between cylinders of the same elementary cell and $\tau_\mathrm{out} = \tau(3a-2R)$ between cylinders of different elementary cells (see SM~\cite{Note1} Sec.~II).
Wu and Hu have predicted the opening of a band gap in the spectrum of such a model when $\tau_\mathrm{in} \ne \tau_\mathrm{out}$, corresponding to $R \ne a$ in our case~\cite{wu15,wu16}. 
Figure~\ref{fig1}~(c) indeed shows that the density of states
$\text{DOS}(f) \propto 1- \langle \text{Re} S_{11}(\vec{r}, f) \rangle_{\vec{r}}$
gets depressed around $f \simeq 7.4$~GHz when $R \ne a$.
The spectral range in which the depression takes place is well described by the tight-binding calculation yielding red lines in Fig.~\ref{fig1}~(c). 

Formally replacing the electron spin operator $\hat{\sigma}$ by a pseudospin corresponding to the orbital angular momentum $\hat{\vec{J}} = \hat{J}_z \vec{e}_z$ of the electromagnetic wave in six-cylinder clusters allows for computing topological invariants by analogy with QSHE: the spin Chern number $C_{\text{SC}}$~\cite{Sheng2006Jul} and the spin Bott index $C_{\text{SB}}$~\cite{Huang2018Sep} (see SM~\cite{Note1} Sec.~V).
The results obtained for the tight-binding model on which our experimental system can be mapped are shown in Fig.~\ref{fig1}~(d).
The topological character of the gap for $R > a$ predicted earlier ~\cite{wu15,wu16} is witnessed by $C_{\text{SC}}$ (computed for the infinite lattice) and $C_{\text{SB}}$ (computed for a lattice of finite size) that become different from zero. Note that the values of $C_{\text{SC}}$ and $C_{\text{SB}}^{(\text{PBC})}$ computed using periodic boundary conditions (PBC) coincide exactly, which is in agreement with recent mathematical literature~\cite{toniolo22}.
We also compute the spin Bott index $C_{\text{SB}}^{(\text{OBC})}$ for a lattice with open boundary conditions (OBC) that correspond to the experimental situation.
The result depends on the lattice size and becomes different from zero at $R$ which is slightly larger than $a$, signaling that the finite sample size perturbs topological properties.
This effect attenuates with increasing lattice size and convergence of $C_{\text{SB}}^{(\text{OBC})}$ towards $C_{\text{SC}} = C_{\text{SB}}^{(\text{PBC})}$ is observed.       

\begin{figure}
    \centering
    \includegraphics{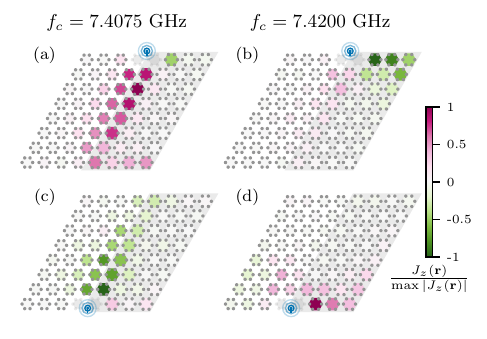}
     \caption{
     Unidirectional propagation of microwaves along edges of a sample
     with a topologically nontrivial band gap.
        The left half of the sample is topologically trivial ($R/a = 0.94$) whereas the right half is topologically nontrivial ($R/a = 1.06$). The color code shows the time-integrated orbital angular momentum of light (pseudospin) in six-cylinder clusters $J_z(\vec{r})$.
        The pulsed source position is indicated by an antenna pictogram near the top or bottom of the boundary between the two halves of the sample.}
     \label{fig:pattern}
\end{figure}

The topological invariants shown in Fig.~\ref{fig1}~(d) are difficult to measure in an experiment.
To observe an experimental manifestation of a topologically nontrivial band gap in our system, we create a sample consisting of two topologically different halves: 
$R < a$, $C_{\text{SC}} = C_{\text{SB}} = 0$ on the left and
$R > a$, $C_{\text{SC}} = C_{\text{SB}} = 1$ on the right.
The interface between the two halves of the sample should carry topologically protected edge states that we evidence by using the transmission coefficients $S_{21}(\vec{r},f)$ to compute the response of the sample to a pulsed excitation (central frequency $f_c$ around 7.4 GHz) by a loop antenna placed at opposite ends of the interface (top or bottom of the interface for the first and second rows of Fig.~\ref{fig:pattern}, respectively).
The spatiotemporal evolution of pseudospin 
$J_z (\vec{r},t) \propto \sum_{n =1}^6
\text{Im} \left[ S_{21}^*(\vec{r} + \Delta\vec{r}_n,t) S_{21}(\vec{r} + \Delta\vec{r}_{n+1},t) \right]$
(see SM~\cite{Note1} Sec.~IV for details) is shown in Supplemental videos~1--4~\footnote{See SM Videos showing the propagation of a Gaussian pulse between topologically distinct regions (videos 1--4) and around a defect (videos 5 and 6).}.
Here $\Delta\vec{r}_n$ is the position of the cylinder $n$ with respect to the center $\vec{r}$ of a six-cylinder cluster.
Figure~\ref{fig:pattern} shows the time integral $J_z(\vec{r})$ of $J_z(\vec{r},t)$.
We clearly see that in all cases, propagation takes place along boundaries between either topologically distinct parts of the sample (Figs.~\ref{fig:pattern}~(a) and (c)) or the topologically nontrivial part and the free space (Figs.~\ref{fig:pattern}~(b) and (d)).
The sign of $J_z$ is locked to the direction of propagation: $J_z < 0$ or $J_z > 0$ for clockwise or counter-clockwise propagation, respectively.
This is in full analogy with QSHE in which electrons with opposite spins $\sigma = \pm 1/2$ propagate in opposite directions along an edge of a 2D sample. 
The direction in which a wave excited by a source at a given position propagates is controlled by the frequency of the source $f_c$: for $f_c$ below (above) the middle of the gap, the system is on the decreasing (rising) part of the dispersion curve of the edge state and the propagation is counter-clockwise (clockwise) with $J_z > 0$ ($J_z < 0$).
This is in contrast with the proposal of Ref.\ \cite{wu15} where the use of a pseudospin-polarized source has been suggested to break the symmetry between modes with positive and negative $J_z$ whereas our source is equally efficient in exciting both.
The fact that a single edge mode dominates in all cases can be explained by an accurate analysis of the band diagram in Fig.\ \ref{fig1}~(b), see SM~\cite{Note1} Sec.~II.C.

\begin{figure}
    \centering
    \includegraphics{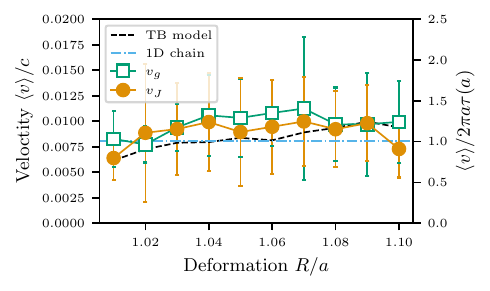}
     \caption{
     Velocity of propagation of the orbital angular momentum $v_J$ (solid orange circles) and group velocity $v_g$ (open green squares) along the interface between topologically distinct parts of the sample compared to the theoretical predictions for the group velocity of the corresponding tight-binding model (black dashed line), and a 1D chain of coupled harmonic oscillators (blue dash-dotted line).
     Six-cylinder cluster size is $R$ and $2a-R$ for the two topologically distinct parts of the sample, respectively.
     Data are averaged over several central frequencies $f_c$ of the source within the band gap.}
     \label{fig:velocity}
\end{figure}

In addition to yielding appealing visual representations of the dynamics of helical edge states (see Supplemental Videos~1--4~\cite{Note2}), our measurements allow for measuring their velocities. We define the group velocity $v_g$ as the velocity of propagation of the maximum of intensity
$I(\vec{r},t) = \sum_{n=1}^6 \left| S_{21}(\vec{r} + \Delta \vec{r}_n, t) \right|^2$.
$v_g$ is shown by open green squares in Fig.~\ref{fig:velocity} and agrees well with the theoretical prediction following from the simulation of the equivalent tight-binding model (black dashed line, see SM~\cite{Note1} Sec.~II.D).
We also define the velocity of pseudospin transport $v_J = [dt_b(s \vec{e}_{\text{edge}})/ds]^{-1}$, where $t_b(\vec{r})$ is the barycenter of $J_z(\vec{r},t)$ in a six-cylinder cluster centered at $\vec{r}$:
$\int_{-\infty}^{t_b(\vec{r})} |J_z(\vec{r},t)| dt = \int_{t_b(\vec{r})}^{\infty} |J_z(\vec{r},t)| dt$, and $\vec{e}_{\text{edge}}$ is a unit vector parallel to the interface between topologically distinct parts of the sample.
In principle, $v_J$ may be different from $v_g$, which would signal that speeds of pseudospin and energy transports are not the same.
However, our measurements yield $v_J \simeq v_g$ within experimental accuracy, see Fig.~\ref{fig:velocity}.
$v_g$ and $v_J$ exhibit very little dependence on $R/a$, remaining roughly constant within the range $R/a = 1.01$---$1.1$ explored in the experiment. 
Furthermore, we observe that the measured velocities are of the same order of magnitude as the velocity predicted and observed in a 1D chain of coupled resonators, i.e. $v_g 
 \approx 2\pi a\tau(a)$  with $a=10$~mm and $\tau(a) = 38.58$~MHz (blue dash-dotted line in Fig.~\ref{fig:velocity}, see SM~\cite{Note1} Sec.~III).
The same calculation done for coupled harmonic oscillators arranged in a honeycomb lattice yields velocities that are a factor of 2 larger (see SM~\cite{Note1} Sec.~III.B).
This observation emphasizes the 1D character of the helical edge states.

\begin{figure}
    \centering
    \includegraphics{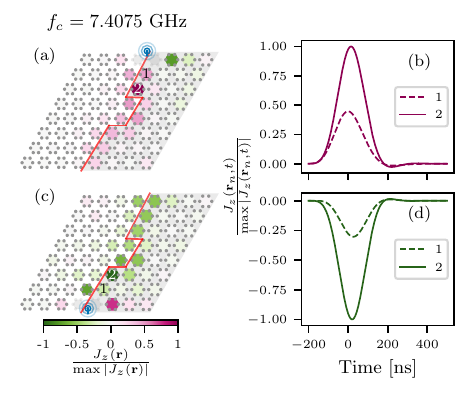}
     \caption{
          Topological protection of edge states against disorder.
          (a) Color-scale plot of the time-integrated pseudospin $J_z( \vec{r})$ for excitation by a loop antenna located at the top of the sample.
          A defect consisting of two six-cylinder clusters of topologically trivial type ($R/a = 0.94$, same as in the topologically trivial left half of the sample) is protruding into the topologically nontrivial right half of the sample where $R/a = 1.06$.
          (b) Temporal profiles of $J_z( \vec{r},t)$ for clusters marked as 1 and 2 in panel (a).
          Panels (c) and (d) are the same as panels (a) and (b) except for the emitting antenna at the bottom of the sample.}
     \label{fig:defect}
\end{figure}

Robustness against scattering is a distinctive feature of edge states arising due to the topologically nontrivial band structure in the bulk \cite{oh13,bernevig13}.
In order to demonstrate it in our system, we prepare a sample with a defect located along the boundary separating the two topologically distinct parts of the sample.
The defect consists of two six-cylinder clusters of size $
R < a$ protruding into the part of the sample where $R > a$, see Fig.~\ref{fig:defect}.
Repeating the measurements reported in Fig.~\ref{fig:pattern}~(a) and (c) produces results shown in Fig.~\ref{fig:defect} and Supplemental Videos 5 and 6~\cite{Note2}.
As we see in Figs.~\ref{fig:defect}~(a) and (c), the defect does not perturb the unidirectional transport of pseudospin.
Microwaves go around the defect without loosing their pseudospin polarization.
Figures~\ref{fig:defect}~(b) and (d) demonstrate the absence of backscattering of microwaves by the defect: the temporal profiles of $J_z(\vec{r},t)$ at the two six-cylinder clusters in front of the defect exhibit single maxima corresponding to the passage of the microwave pulse but do not show any secondary peak that might be caused by reflection of the wave by the defect
(see also SM \cite{Note1} Sec.~VI). 

Topologically protected edge states in TR-symmetric photonic systems have been proposed as an efficient means to guide light in optoelectronic devices \cite{khanikaev13,chen14,ma15,yang18}. 
The low velocity of propagation of pseudospin-polarized signal measured in our experiments $v = v_J \simeq v_g \sim 3 \times 10^6 \text{~m/s} \sim c/100$ imposes constraints on the information transfer rate of an optical communication channel using such states even though the precise value of $v$ can be adjusted according to the relation $v \approx 2\pi a \tau(a)$, which allows for tuning $v$ between $\sim c/50$ and $\sim c/1200$ by varying the nearest-neighbor spacing $a$ in our experiment from 7 to 20~mm. Another possible application of topological edge states is their use as logical qubits that would be resilient to noise due to their nonlocal character \cite{lukin01,bluvstein23}.
In this case, the finite speed of signal propagation limits the maximum rate at which such a qubit can be addressed because of nonzero time needed for the qubit to reach steady state.
For a 2D logical qubit composed of $10 \times 10$ elementary cells (a smaller system may not exhibit a proper band gap in the bulk), the time for a signal to go around the system  is $\sim 40 \times 3a/v = 0.2$--5~µs. This is already much longer than the 25~ns needed to operate a single-qubit gate in the contemporary superconducting quantum processors \cite{arute19}. The above examples show that the dynamical transport properties of  topologically protected edge states may be crucial for the design of their practical applications.

\begin{acknowledgments}
This work was funded by the Agence Nationale de la Recherche (Grant No. ANR-20-CE30-0003 LOLITOP) and supported by the French government through the France 2030 investment plan managed by the Agence Nationale de la Recherche as part of the Initiative of Excellence Universit\'{e} C\^{o}te d’Azur under Reference No. ANR-15-IDEX-01. The authors are grateful to the Universit\'{e} C\^{o}te d’Azur’s Center for High-Performance Computing (OPAL infrastructure) for providing resources and support.
\end{acknowledgments}

\bibliography{refstopo}

\ifarXiv
    \hfill
    

\section*{Supplementary material}

# Supplemental Material for "Dynamics of transport by helical edge states"

Luis Alberto Razo López,[1] Pierre Wulles,[2] Geoffroy J. Aubry,[1, *] Sergey E. Skipetrov,[2, †] and Fabrice Mortessagne[1, ‡]

[1]*Université Côte d'Azur, CNRS, Institut de Physique de Nice (INPHYNI), France*
[2]*Université Grenoble Alpes, CNRS, LPMMC, Grenoble, France*

## CONTENTS

---

* geoffroy.aubry@cnrs.fr
† sergey.skipetrov@lpmmc.cnrs.fr
‡ fabrice.mortessagne@univ-cotedazur.fr

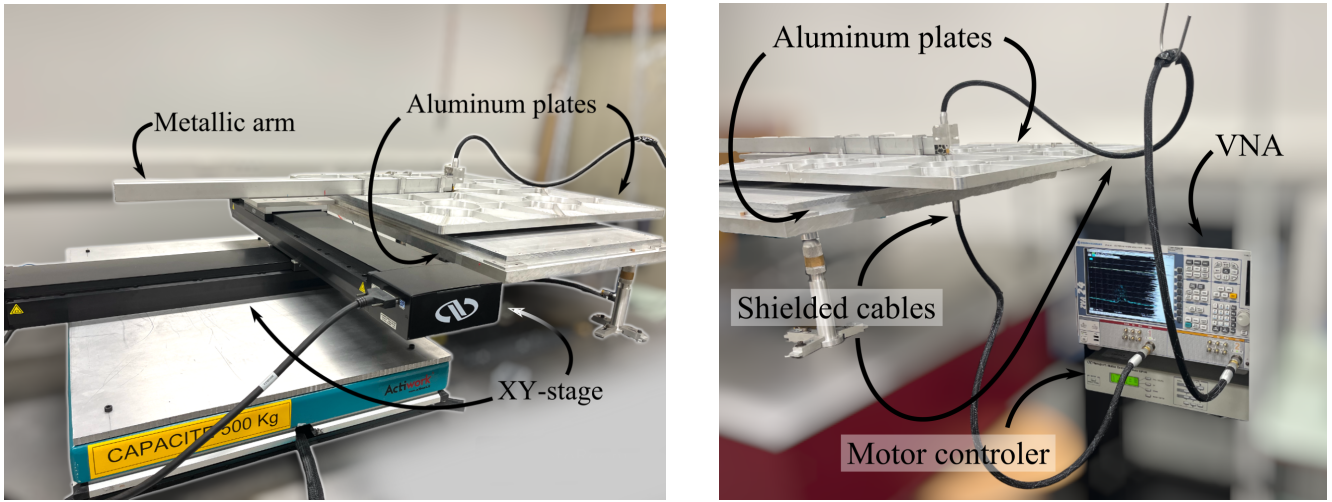

FIG. S1. Photographs of the experimental platform and its elements.

## I. DETAILS OF THE EXPERIMENTAL SETUP

### A. The electromagnetic platform

All the experiments are conducted using a quasi-two-dimensional microwave cavity made of two parallel aluminum plates placed at $z = 0$ and $z = h = 13$ mm. The top plate is suspended on a horizontal metallic arm attached to a motorized XY-stage (Newport IMS600C and Newport ESP301) and movable in the $xy$ plane. The cavity is open along both $x$ and $y$ axes. The XY-stage provides a 200 nm precision in each direction. Electromagnetic waves are excited and measured using two antennas that penetrate into the cavity through two holes drilled at the centers of top (antenna 1) and bottom (antenna 2) plates, respectively. The antennas are connected to a two-port vector network analyzer (VNA, Rohde & Schwarz ZVA 24) via two shielded flexible coaxial cables (Flexco NTC195-50R49). Figure S1 illustrates the experimental platform.

VNA operates in a frequency range from 10 MHz to 24 GHz corresponding to wavelengths in vacuum from 1.25 cm to 29 m. As VNA measures modifications in amplitude and phase of a reference signal, it provides a $2 \times 2$ complex scattering matrix $\hat{S}$ relating input signals $V_{\text{in}}^1$, $V_{\text{in}}^2$ to the measured signals $V_{\text{out}}^1$, $V_{\text{out}}^2$ of the two antennas:

$$\begin{pmatrix} S_{11} & S_{12} \\ S_{21} & S_{22} \end{pmatrix} \begin{pmatrix} V_{\text{in}}^1 \\ V_{\text{in}}^2 \end{pmatrix} = \begin{pmatrix} V_{\text{out}}^1 \\ V_{\text{out}}^2 \end{pmatrix} \tag{S1}$$

In this work, we analyze the reflection coefficient $r = S_{11}$ of the movable top antenna 1 and the transmission coefficient $t = S_{21}$ from the movable top antenna 1 to the immobile bottom antenna 2.

### B. Microwave antennas

Our experiments are carried out for TE polarization of electromagnetic waves that are selectively excited in the experimental setup via the specific geometry of the antennas, see Fig. S2(a). As shown in the figure, a wire is folded into a loop in the $xy$ plane and then the circuit is closed by welding the end of the wire to the beginning of the loop.

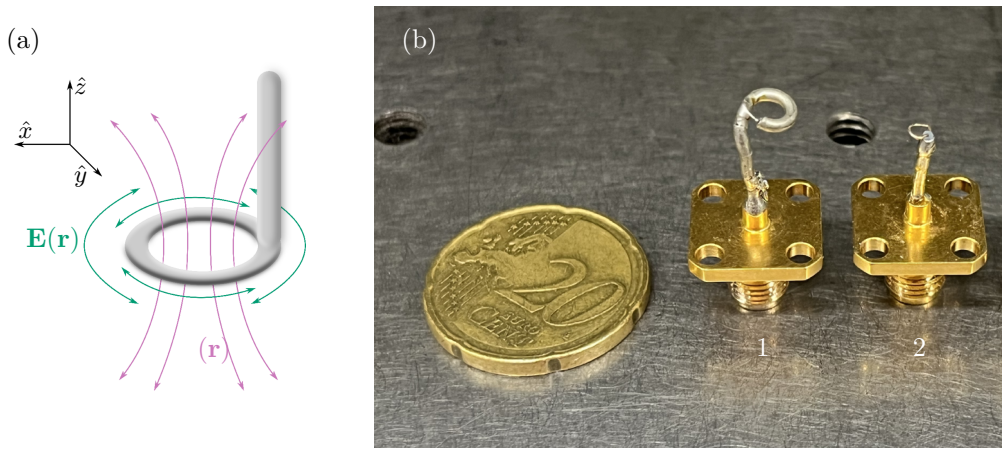

FIG. S2. (a) Diagram of the electric and magnetic field radiated by a loop antenna. (b) Photograph showing the loop antennas used in our experiments and a coin for scale.

An oscillating current in the loop excites TE electromagnetic field with $E_z = 0$ and $H_z \propto E_\varphi$.

Two loop antennas used in our experiments are shown in Fig. S2(b). They have loop diameters of 2 and 5.5 mm, respectively. The antennas exhibit measurable resonances produced by standing waves inside them. These resonances are avoided in the experiment by working only in certain (usable) frequency ranges. The small and big loop antennas have usable frequency ranges between 7–10 GHz and 5–8 GHz, respectively.

### C. Dielectric cylinders

The flexibility of our experimental setup in mainly due to the fact that we can locally modify the permittivity of the microwave cavity by placing dielectric cylinders inside it. In this work, we use ceramic cylinders made of TiZrNbZnO (Exxelia Temex manufacturer serie E6000), $h_c = 5$ mm in height and $r_c = 3$ mm in radius. They are characterized by a high relative permittivity $\epsilon_c = 45$, no magnetic properties (relative permeability $\mu_c = 1$, refractive index $n_c = \sqrt{\mu_c \epsilon_c} = \sqrt{\epsilon_c} \approx 7$) and a large quality factor $Q \approx 8000$ at 5 GHz. Figure S3(a) shows a photograph of an actual dielectric cylinder used in our experiments.

The dielectric cylinders are precisely placed on the bottom plate at predefined locations by dropping them through a metallic tube. In order to use the motorized XY-stage to place cylinders, the top plate is removed and replaced by the tube. Figure S3(b) shows an image of the cylindrical metallic tube used in the experiment to place the dielectric cylinders.

## II. MAPPING OF THE EXPERIMENTAL SYSTEM TO THE TIGHT-BINDING MODEL

### A. TE polarized waves in the empty cavity

The empty cavity can be modeled by Maxwell's equations without sources [S1]. Considering a harmonic monochromatic time dependence with radial frequency $\omega = 2\pi f$: $\mathbf{E}(\mathbf{r}, t) = \mathbf{E}(\mathbf{r})e^{-i\omega t}$, $\mathbf{H}(\mathbf{r}, t) = \mathbf{H}(\mathbf{r})e^{-i\omega t}$, we obtain Helmholtz

equations for **E** and **H**:

$$\Delta \mathbf{E}(\mathbf{r}) = -k^2 \mathbf{E}(\mathbf{r}), \qquad \Delta \mathbf{H}(\mathbf{r}) = -k^2 \mathbf{H}(\mathbf{r}) \tag{S2}$$

where $c$ is the speed of light in free space and $k = \omega/c$ the wave number. Boundary conditions at the conducting plates are [S1]

$$\mathbf{e}_z \times \mathbf{E}(\mathbf{r})\Big|_{z=0,h} = 0, \quad \mathbf{e}_z \cdot \mathbf{H}(\mathbf{r})\Big|_{z=0,h} = 0 \tag{S3}$$

For TE polarized waves, we have $E_z(\mathbf{r}) = 0$ and the boundary condition for $H_z$ is

$$H_z(\mathbf{r})\Big|_{z=0,h} = 0 \tag{S4}$$

The boundary condition (S4) imposes $z$-dependence of $H_z$: $H_z \propto \sin(k_z z)$ with $k_z = j\pi/h$ ($j = 1, 2, \ldots$). The condition $k_z \leq k$ gives cutoff frequencies of the propagating modes in the cavity:

$$f_{j,\text{cut}} = j\frac{c}{2h} \qquad (j = 1, 2, \ldots). \tag{S5}$$

Thus, no propagating TE electromagnetic waves exist in the cavity at frequencies $f < f_{1,\text{cut}}$.

### B. TE polarized waves in the presence of dielectric cylinders

Our experiments are performed in a frequency range $c/2n_c h_c < f < c/2h$ corresponding to $f = 4.7$–$11.5$ GHz. In this range, TE electromagnetic waves can freely propagate inside a dielectric cylinder but exhibit evanescent behavior in the air surrounding it. As a result, for a single cylinder placed between the conducting plates, the electromagnetic field is confined inside the cylinder leading to sharp well-isolated resonances at frequencies depending on the separation $h$ between the plates [S2]. We work around the first of these resonances and determine its parameters (frequency $f_0$ and width $\delta f_0$) by fitting the reflection signal $S_{11}$ with a Lorentzian function (see e.g. Fig. S4(a) and (b))

$$S_{11}(f) = \frac{A_0}{f - f_0 + i\delta f_0} \tag{S6}$$

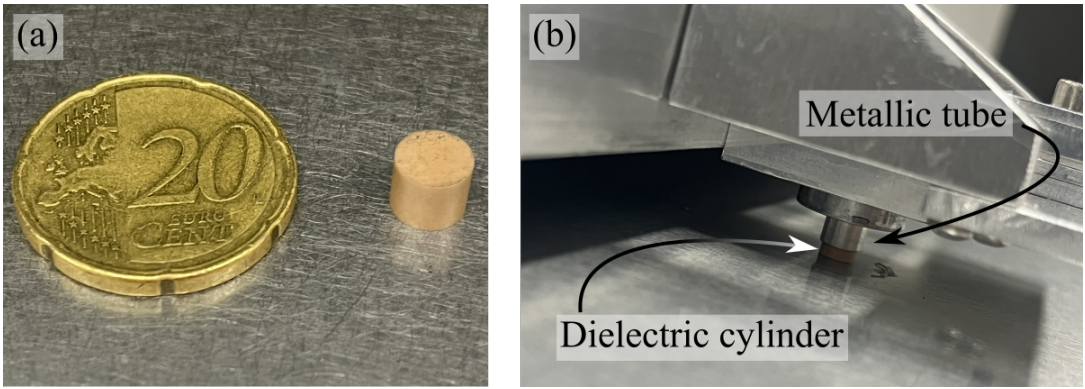

FIG. S3. (a) Photograph of a dielectric cylinder near a coin for scale. (b) Photograph of the metallic tube used to place the cylinders.

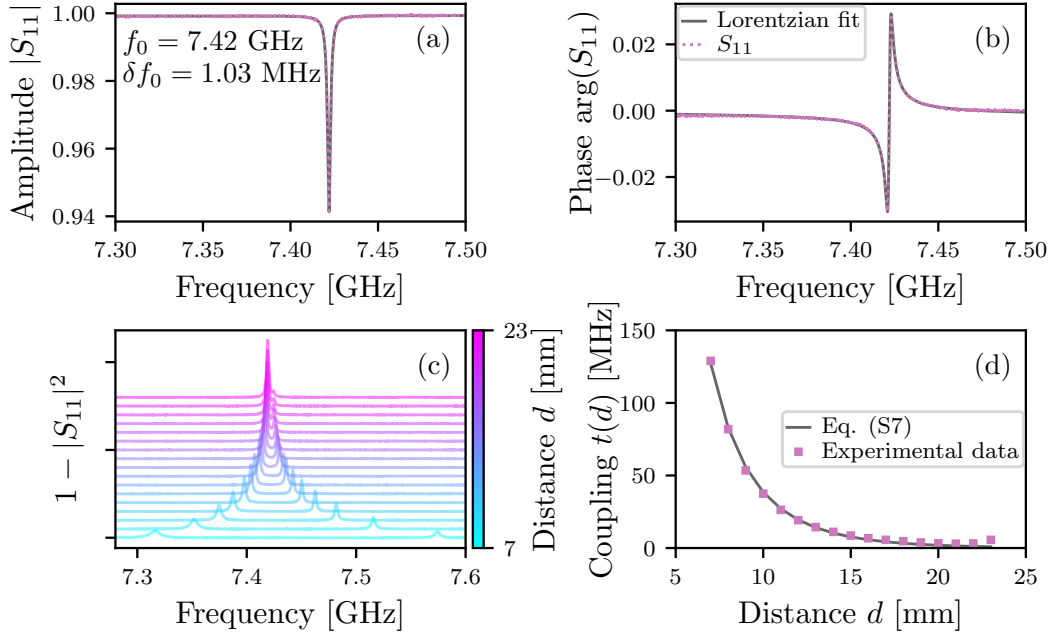

FIG. S4. (a) Amplitude and (b) phase of the reflection coefficient $S_{11}$ for a single cylinder in a cavity of plate spacing $h = 13$ mm (pink points). A fit of Eq. (S6) (gray line) allows to find the resonance parameters. (c) $1 - |S_{11}(f)|^2$ for two cylinders separated by a distance $d$. The antenna is placed at a distance of 8 mm from one of the cylinders. (d) Resonance splitting as a function of the distance between two resonators (squares). A fit of Eq. (S7) is shown by the solid line and yields $\kappa = 43.58$ MHz and $\gamma_1 = 0.1985$ mm$^{-1}$.

Consider now two identical cylinders with resonant frequencies $f_0$ separated by a distance $d$. Coupling between cylinders results in a symmetric splitting $\Delta f_0(d)$ of the single-cylinder resonance. Figure S4(c) depicts $1 - |S_{11}(f)|^2$ for two coupled cylinders at different distances between their centers $d$. A fit of $S_{11}$ by a sum of two Lorentzian functions (S6) allows for extracting the frequency splitting $\Delta f_0$ as a function of the distance $d$. Defining a coupling parameter [S2–S4]

$$t(d) = \frac{\Delta f_0}{2} \approx \kappa K_0\left(\gamma_1 \frac{d}{2}\right)\left[K_2\left(\gamma_1 \frac{d}{2}\right) + K_0\left(\gamma_1 \frac{d}{2}\right)\right] \tag{S7}$$

where $K_n$ is the modified Bessel function of order $n$ and $\kappa$, $\gamma_1$ are parameters determined from the fit in Fig. S4(d), allows us to model the system of two cylinders in the cavity by a Hamiltonian matrix

$$\hat{H}(d) = \begin{pmatrix} f_0 & t(d) \\ t(d) & f_0 \end{pmatrix} \tag{S8}$$

By extending the Hamiltonian (S8) to $N$ cylinders, we obtain a tight-binding model in which any two cylinders $m$ and $n$ are coupled by a coefficient $t(d_{mn})$ depending on the distance $d_{mn}$ between their centers according to Eq. (S7). We use this Hamiltonian to model our experimental setup and to obtain results shown in Figs. 1(b), 1(c)(red lines), 1(d), 3(black dashed line).

### C. Band diagram and edge states

To obtain a theoretical interpretation of our experimental results concerning edge modes, we consider an infinite ribbon with the same types of boundaries between topologically distinct parts of the sample and the sample and the air as in our experimental setup, see the inset of Fig. 1(b) of the main text. Its typical band diagram is presented in Fig. 1(b). It is calculated in the tight-binding approximation introduced in Sec. II B and assuming that the left half of the ribbon is topologically trivial ($R = 0.94a$) whereas the right one has a topological band gap ($R = 1.06a$). The edge mode arising at the boundary between topologically distinct parts of the sample is shown in yellow. The blue line represents the edge mode arising at the boundary between the topological half of the ribbon and the air. Let us now establish a correspondence between this band diagram and edge modes shown in Fig. 2. The first column of Fig. 2 is obtained for a frequency $f_c$ that is only slightly above the minimum of the upper yellow band in Fig. 1(b). At this frequency, only "yellow" edge states exist and we clearly observe propagation of pseudospin-polarized optical currents along the boundary between two topologically distinct halves of the sample in Figs. 2(a) and (c). No "blue" modes exist at this frequency and therefore an antenna cannot emit along the boundary between the sample and the air. Weak signals still observed along this boundary in Figs. 2(a) and (c) are due to the finite bandwidth of the exciting pulse and the finite lifetime (and hence nonzero spectral width) of modes. As a consequence, our previous reasoning implicitly assuming monochromatic excitation and infinite-lifetime modes, applies only approximately. In a real experiment, an antenna excites all modes whatever the central frequency $f_c$ of the emitted pulse. However, the amplitude of excitation varies depending on the mode frequency $f$, with much stronger signals in modes with $f$ closest to $f_c$.

At a higher frequency $f_c$ corresponding to the second column of Fig. 2 the situation is more involved. This frequency roughly corresponds to the minimum of the upper blue band in Fig. 1(b) but it matches the yellow band as well. Thus, at this frequency edge states exist at both boundaries. However, now most of the emission goes to the blue mode as we clearly observe in Figs. 2(b) and (d). We explain this by the fact that the power emitted by an antenna into a given mode should be roughly proportional to DOS associated to this mode. In its turn, DOS $\propto 1/v_g$. Because the group velocity $v_g$ of the blue mode is smaller than that of the yellow mode for $f_c$ corresponding to the second column of Fig. 2, the antenna emits most of the power into the blue mode. We consistently observe preferable emission into the blue mode at all frequencies inside the band gap at which both modes exist. This suggests that additional factors may break the symmetry between the modes and privilege the blue mode, such as a better impedance matching, for example, or the fact that the spectrum of our experimental setup is discrete (due to the finite sample size), in contrast to the spectrum of an infinite ribbon shown in Fig. 1(b).

### D. Calculation of the group velocity

The theoretical prediction of the group velocity shown in the main text (dashed line in Fig. 3) is made by reconstructing the band diagram of the system using the PythTB python package [S5] and Eq. (S7) with the corresponding parameters $\kappa = 43.58$ MHz, $\gamma_1 = 0.1985$ mm$^{-1}$ and $f_0 = 7.42$ GHz [see Fig. S5(a)]. The simulated samples have the same spatial configuration as the experimental ones: topologically trivial ($R < a$) left half of the sample and topologically nontrivial ($R > a$) right half of the sample. Additionally, we impose an effective coupling cutoff between

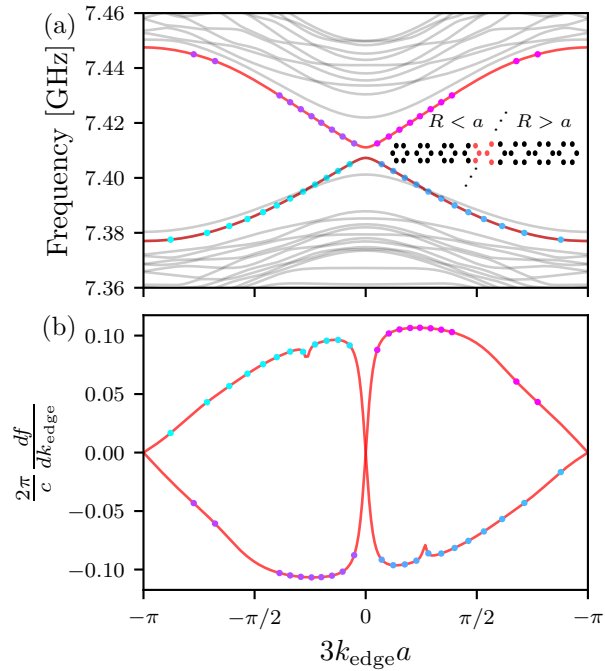

FIG. S5. (a) Example of band structure of an infinite ribbon (see inset for a sketch of its single unit cell repeated along dotted lines) divided in two halves with $R < a$ on the left and $R > a$ on the right, and having the same types of boundaries as our sample, in the tight-binding approximation. Red lines show bands corresponding to the edge states at the interface between the two halves of the sample. (b) Derivative of the bands corresponding to the edge states at the interface with respect to the wave number $k_{\text{edge}}$. Colored dots indicate the frequencies where helical propagation is detected in the experiment.

cylinders separated by more than 30 mm. The group velocity of the modes at the interface between two topologically distinct halves of the sample is computed as a derivative of their frequency bands: $v_g = 2\pi df/dk_{\text{edge}}$ where the wave vector $\mathbf{k}_{\text{edge}}$ is parallel to the interface, with a precision of $\Delta k_{\text{edge}} = 2\pi/501$ [see Fig. S5(b)]. Finally, results shown in Fig. 3 by the dashed line are obtained by averaging $|v_g|$ over frequencies at which helical mode propagation is experimentally detected (see colored dots in Fig. S5). Good agreement with the measurements is observed.

### III. GROUP VELOCITY IN ARRAYS OF COUPLED OSCILLATORS

To prove the 1D nature of the helical edge modes, in the present section we compute group velocities in a 1D linear chain and a 2D honeycomb lattice of coupled resonators assuming that they can be modeled by corresponding arrays of coupled linear oscillators. We compare the calculated values with those measured in experiments conducted with 1D chains of dielectric cylinders.

## A. 1D linear chain

Consider a 1D linear chain of coupled harmonic oscillators with damping described by

$$\ddot{x}_n = -\omega_0^2 x_n - 2\zeta\omega_0 \dot{x}_n + 2\tilde{t}\omega_0^2(x_{n+1} - 2x_n + x_{n-1}) \tag{S9}$$

where $\omega_0 = 2\pi f_0$ is the resonance angular frequency of oscillators, $\zeta$ the damping ratio, and $\tilde{t}$ the coupling between the resonators. We look for plane waves solutions of this equation $x_n = X_0 \exp(-i(\omega t - kna))$ with $X_0$ a complex number, $\omega$ a (real) angular frequency, and $k = k' + ik''$ a (complex) wavenumber. $a$ is the distance between two neighboring oscillators. Substituting

$$\dot{x}_n = -i\omega x_n \tag{S10}$$

$$\ddot{x}_n = -\omega^2 x_n \tag{S11}$$

$$x_{n\pm 1} = \exp(\pm ika)x_n = \exp(\pm ik'a)\exp(\mp k''a)x_n \tag{S12}$$

into Eq. (S9), we obtain

$$-\omega^2 = -\omega_0^2 + 2i\zeta\omega_0\omega + 2\tilde{t}\omega_0^2(\exp(ik'a)\exp(-k''a) - 2 + \exp(-ik'a)\exp(k''a)) \tag{S13}$$

Let $\Omega = \omega/\omega_0$. Then

$$-\Omega^2 = -1 + 2i\zeta\Omega + 2\tilde{t}(\exp(ik'a)\exp(-k''a) - 2 + \exp(-ik'a)\exp(k''a)) \tag{S14}$$

We separate real and imaginary parts of Eq. (S14):

$$-\Omega^2 = -1 + 2\tilde{t}(\cos(k'a)\exp(-k''a) - 2 + \cos(-k'a)\exp(k''a)) \tag{S15}$$

$$0 = 2\zeta\Omega + 2\tilde{t}(\sin(k'a)\exp(-k''a) + \sin(-ik'a)\exp(k''a)) \tag{S16}$$

which we can rewrite as

$$0 = \Omega^2 - 1 + 4\tilde{t}\cos(k'a)\cosh(k''a) - 4\tilde{t} \tag{S17}$$

$$0 = 2\zeta\Omega - 4\tilde{t}\sin(k'a)\sinh(k''a) \tag{S18}$$

Assume that damping is weak and thus $k''a \ll 1$. Expanding $\cosh(k''a)$ and $\sinh(k''a)$ around $k'' = 0$, we obtain

$$0 = \Omega^2 - 1 + 4\tilde{t}\cos(k'a)\left(1 + \frac{(k''a)^2}{2}\right) - 4\tilde{t} \tag{S19}$$

$$0 = 2\zeta\Omega - 4\tilde{t}\sin(k'a)k''a \tag{S20}$$

At the first order in $k''a$, Eq. (S19) becomes

$$1 - \Omega^2 + 4\tilde{t} = 4\tilde{t}\cos(k'a) \tag{S21}$$

From Eq. (S21), we directly get the dispersion relation

$$\Omega = \sqrt{1 + 4\tilde{t}(1 - \cos(k'a))} \tag{S22}$$

and the group velocity

$$\frac{\partial \Omega}{\partial k'a} = \frac{2\tilde{t}\sin(k'a)}{\sqrt{1 + 4\tilde{t}(1-\cos(k'a))}} \tag{S23}$$

If the coupling is small $\tilde{t} \ll 1$, $\frac{\partial \Omega}{\partial k'a} \simeq 2\tilde{t}\sin(k'a)$ and therefore averaged over the first Brillouin zone, $\langle|\frac{\partial \Omega}{\partial k'a}|\rangle \simeq \tilde{t}$. Going back to dimensional units, we have $\frac{\partial \omega}{\partial k'} = a\omega_0 \frac{\partial \Omega}{\partial k'a}$, and therefore the group velocity is of the order of $a\omega_0 \tilde{t} = 2\pi a t(a)$. Using $a = 1$ cm as in the experiment, as well as Eq. (S7) for $t(a)$, we obtain the average group velocity of $v_g \simeq 2\pi a t(a) \simeq 2.4 \cdot 10^6$ m/s $\simeq 0.8 \cdot 10^{-2} c$. We measured the propagation velocity of microwaves in 1D a linear chain of resonators and found velocities of the order of $10^{-2} c$, consistent with this calculation.

### B. 2D honeycomb lattice

A honeycomb lattice is a triangular lattice [primitive vectors $\mathbf{a} = a\mathbf{e}_x$, $\mathbf{b} = a(\frac{1}{2}\mathbf{e}_x + \frac{\sqrt{3}}{2}\mathbf{e}_y)$] with two atoms per unit cells at $\bullet = (0,0)$ and $\bullet = (0, \frac{a}{2}\mathbf{e}_y)$. The Bravais lattice is $\mathbf{r}_{n,m} = n\mathbf{a} + m\mathbf{b}$ with $n$ and $m$ two integers. Because we have two atoms per unit cell, we have two coupled equations instead of Eq. (S9):

$$\begin{cases} \ddot{z}_{n,m,\bullet} = -\omega_0^2 z_{n,m,\bullet} + 2\omega_0^2 \tilde{t}(-3 z_{n,m,\bullet} + z_{n,m,\bullet} + z_{n,m-1,\bullet} + z_{n+1,m-1,\bullet}) \\ \ddot{z}_{n,m,\bullet} = -\omega_0^2 z_{n,m,\bullet} + 2\omega_0^2 \tilde{t}(-3 z_{n,m,\bullet} + z_{n,m,\bullet} + z_{n,m+1,\bullet} + z_{n-1,m+1,\bullet}) \end{cases} \tag{S24}$$

Here we neglect damping as we have seen in Sec. III A that weak damping have little influence on the dispersion relation. We look for plane-wave solutions of Eq. (S24):

$$\begin{pmatrix} z_{n,m,\bullet} \\ z_{n,m,\bullet} \end{pmatrix} = \begin{pmatrix} Z_{0,\bullet} \\ Z_{0,\bullet} \end{pmatrix} \exp[-i(\omega t - \mathbf{k} \cdot (n\mathbf{a} + m\mathbf{b}))] \tag{S25}$$

with $Z_{0,\bullet}, Z_{0,\bullet}$ two complex numbers, $\omega$ the (real) angular frequency, and $\mathbf{k} = k_x \mathbf{e}_x + k_y \mathbf{e}_y = \mathbf{k}' + i\mathbf{k}'' = (k_x' + ik_x'')\mathbf{e}_x + (k_y' + ik_y'')\mathbf{e}_y$ the (complex) wave number. Because we neglected damping, we set $\mathbf{k}'' = 0$. Using

$$\dot{z}_{n,m,\bullet/\bullet} = -i\omega z_{n,m,\bullet/\bullet} \tag{S26}$$

$$\ddot{z}_{n,m,\bullet/\bullet} = -\omega^2 z_{n,m,\bullet/\bullet} \tag{S27}$$

$$z_{n\pm 1,m,\bullet/\bullet} = e^{\pm i\mathbf{k}\cdot\mathbf{a}} z_{n,m,\bullet/\bullet} \tag{S28}$$

$$z_{n,m\pm 1,\bullet/\bullet} = e^{\pm i\mathbf{k}\cdot\mathbf{b}} z_{n,m,\bullet/\bullet} \tag{S29}$$

we can rewrite Eq. (S24) as

$$\begin{cases} -\omega^2 z_{n,m,\bullet} = -\omega_0^2 z_{n,m,\bullet} \\ \qquad\qquad + 2\omega_0^2 \tilde{t}\big(-3 z_{n,m,\bullet} + z_{n,m,\bullet} + e^{-i\mathbf{k}\cdot\mathbf{b}} z_{n,m,\bullet} + e^{i\mathbf{k}\cdot(\mathbf{a}-\mathbf{b})} z_{n,m,\bullet}\big) \\ -\omega^2 z_{n,m,\bullet} = -\omega_0^2 z_{n,m,\bullet} \\ \qquad\qquad + 2\omega_0^2 \tilde{t}\big(-3 z_{n,m,\bullet} + z_{n,m,\bullet} + e^{i\mathbf{k}\cdot\mathbf{b}} z_{n,m,\bullet} + e^{i\mathbf{k}\cdot(-\mathbf{a}+\mathbf{b})} z_{n,m,\bullet}\big) \end{cases} \tag{S30}$$

or in a matrix form

$$\hat{M} \begin{pmatrix} z_{n,m,\bullet} \\ z_{n,m,\bullet} \end{pmatrix} = \begin{pmatrix} 0 \\ 0 \end{pmatrix} \tag{S31}$$

with

$$\hat{M} = \begin{bmatrix} \omega^2 - \omega_0^2 - 6\omega_0^2\tilde{t} & 2\omega_0^2\tilde{t}\left(1 + e^{-i\mathbf{k}\cdot\mathbf{b}} + e^{i\mathbf{k}\cdot(\mathbf{a}-\mathbf{b})}\right) \\ 2\omega_0^2\tilde{t}\left(1 + e^{i\mathbf{k}\cdot\mathbf{b}} + e^{i\mathbf{k}\cdot(-\mathbf{a}+\mathbf{b})}\right) & \omega^2 - \omega_0^2 - 6\omega_0^2\tilde{t} \end{bmatrix} \tag{S32}$$

Equation (S31) has nontrivial solutions only if $\det \hat{M} = 0$ or

$$\left(\omega^2 - \omega_0^2 - 6\omega_0^2\tilde{t}\right)^2 - 4\omega_0^4\tilde{t}^2\left|1 + e^{-i\mathbf{k}\cdot\mathbf{b}} + e^{i\mathbf{k}\cdot(\mathbf{a}-\mathbf{b})}\right|^2 = 0 \tag{S33}$$

Let $\alpha = 1 + e^{-i\mathbf{k}\cdot\mathbf{b}} + e^{i\mathbf{k}\cdot(\mathbf{a}-\mathbf{b})}$. We then have

$$\left(\omega^2 - \omega_0^2(1 + 6\tilde{t})\right)^2 - 4\omega_0^4\tilde{t}^2|\alpha|^2 = 0 \tag{S34}$$

and finally

$$\omega^4 - 2\omega_0^2(1 + 6\tilde{t})\omega^2 + \omega_0^4(1 + 6\tilde{t})^2 - 4\omega_0^4\tilde{t}^2|\alpha|^2 = 0 \tag{S35}$$

This is a quadratic equation with respect to $\omega^2$. Its determinant is

$$\Delta = 4\omega_0^4(1 + 6\tilde{t})^2 - 4\omega_0^4(1 + 6\tilde{t})^2 + 16\omega_0^4\tilde{t}^2|\alpha|^2 = 16\omega_0^4\tilde{t}^2|\alpha|^2 \tag{S36}$$

and its solutions are

$$\omega^2 = \omega_0^2(1 + 6\tilde{t}) \pm 2\omega_0^2\tilde{t}|\alpha| = \omega_0^2\left[1 + 2\tilde{t}(3 \pm |\alpha|)\right] \tag{S37}$$

Thus

$$\Omega = \frac{\omega}{\omega_0} = \sqrt{1 + 2\tilde{t}(3 \pm |\alpha|)}. \tag{S38}$$

This dispersion relation can be differentiated to obtain the dimensionless group velocity:

$$\frac{\partial \Omega}{\partial \mathbf{k}'a} = \frac{\partial \Omega}{\partial k'_x a}\mathbf{e}_x + \frac{\partial \Omega}{\partial k'_y a}\mathbf{e}_y \tag{S39}$$

and

$$\left|\frac{\partial \Omega}{\partial \mathbf{k}'a}\right| = \sqrt{\left(\frac{\partial \Omega}{\partial k'_x a}\right)^2 + \left(\frac{\partial \Omega}{\partial k'_y a}\right)^2} \tag{S40}$$

Using

$$\frac{\partial \Omega}{\partial k'_x a} = \frac{\pm 2\tilde{t}}{\Omega}\frac{\partial |\alpha|}{\partial k'_x a} \tag{S41}$$

we get

$$\left|\frac{\partial \Omega}{\partial \mathbf{k}'a}\right| = \frac{2\tilde{t}}{\Omega}\sqrt{\left(\frac{\partial |\alpha|}{\partial k'_x a}\right)^2 + \left(\frac{\partial |\alpha|}{\partial k'_y a}\right)^2} \tag{S42}$$

A color-scale plot of dimensionless group velocity is presented Fig. S6. A numerical evaluation yields an average over the first Brillouin zone $\left\langle \frac{1}{2\tilde{t}}\left|\frac{\partial \Omega}{\partial \mathbf{k}'a}\right|\right\rangle \simeq 1.15$. A very similar value is obtained by computing the average in the directions defined by $\mathbf{a}$ or $\mathbf{b}$ which are the directions of propagation of the helical edge states in the experiment presented in the main text. Therefore, going back to dimensional units we obtain $\langle |v_g|\rangle \simeq 1.15 a\omega_0 2\tilde{t} = 4.6\pi at(a)$. This is more than twice the value for the 1D linear chain (see Sec. III A). The agreement of the group velocity measured in our experiments for helical edge modes with the calculation for 1D chain rather than with the one for the 2D lattice, emphasizes the 1D nature of the helical edge modes.

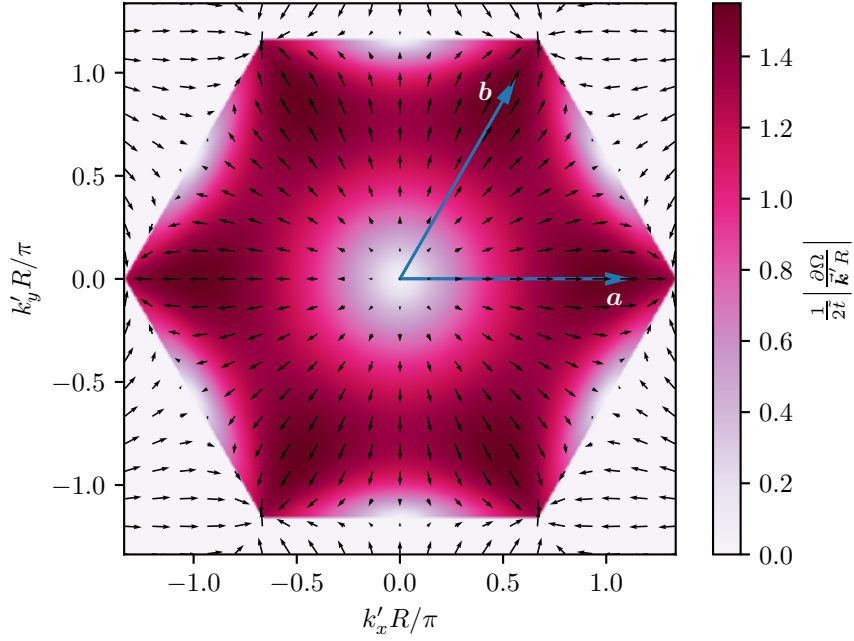

FIG. S6. Group velocity computed for a 2D array of coupled oscillators on a honeycomb lattice. The color encodes the magnitude and the arrows show the direction of velocity.

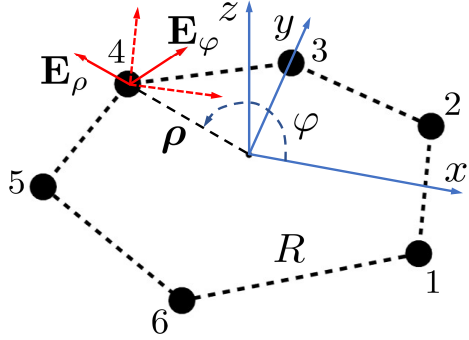

FIG. S7. Hexagonal cluster of six cylinders in a plane $z = $ const.

## IV. ANGULAR MOMENTUM OF LIGHT IN A SIX-CYLINDER CLUSTER

Consider a hexagonal arrangement of six cylinders in a plane $z = $ const, see Fig. S7. For TE modes, the electric field is in the $xy$ plane and the magnetic field is perpendicular to it:

$$\mathbf{E}(\mathbf{r},t) = E_x(\mathbf{r},t)\mathbf{e}_x + E_y(\mathbf{r},t)\mathbf{e}_y = E_\rho(\mathbf{r},t)\mathbf{e}_\rho + E_\varphi(\mathbf{r},t)\mathbf{e}_\varphi \tag{S43}$$

$$\mathbf{H}(\mathbf{r},t) = H_z(\mathbf{r},t)\mathbf{e}_z \tag{S44}$$

where $\{\mathbf{e}_x, \mathbf{e}_y, \mathbf{e}_z\}$ and $\{\mathbf{e}_\rho, \mathbf{e}_\varphi, \mathbf{e}_z\}$ are basis unit vectors of Cartesian and cylindrical reference frames, respectively. According to Maxwell's equations,

$$\frac{\partial \mathbf{E}(\mathbf{r},t)}{\partial t} = \boldsymbol{\nabla} \times \mathbf{H}(\mathbf{r},t) \tag{S45}$$

Assuming monochromatic waves, $\mathbf{E}(\mathbf{r},t) = \mathbf{E}(\mathbf{r})e^{-i\omega t}$, we have

$$\mathbf{E}(\mathbf{r}) = \frac{1}{-i\omega} \boldsymbol{\nabla} \times \mathbf{H}(\mathbf{r}) \tag{S46}$$

where

$$\begin{aligned}
\boldsymbol{\nabla} \times \mathbf{H} &= \left(\frac{\partial H_z}{\partial y} - \frac{\partial H_y}{\partial z}\right)\mathbf{e}_x + \left(\frac{\partial H_x}{\partial z} - \frac{\partial H_z}{\partial x}\right)\mathbf{e}_y + \left(\frac{\partial H_y}{\partial x} - \frac{\partial H_x}{\partial y}\right)\mathbf{e}_z \\
&= \left(\frac{1}{\rho}\frac{\partial H_z}{\partial \varphi} - \frac{\partial H_\varphi}{\partial z}\right)\mathbf{e}_\rho + \left(\frac{\partial H_\rho}{\partial z} - \frac{\partial H_z}{\partial \rho}\right)\mathbf{e}_\varphi + \frac{1}{\rho}\left(\frac{\partial(\rho H_\varphi)}{\partial \rho} - \frac{\partial H_\rho}{\partial \varphi}\right)\mathbf{e}_z
\end{aligned} \tag{S47}$$

We thus have

$$E_\rho = \frac{1}{-i\omega\rho}\frac{\partial H_z}{\partial \varphi}, \quad E_\varphi = \frac{1}{i\omega}\frac{\partial H_z}{\partial \rho} \tag{S48}$$

The energy flux around the 6-cylinder ring is proportional to the Poynting vector

$$\mathbf{S}(\mathbf{r}) = \mathrm{Re}\left[\mathbf{E}(\mathbf{r}) \times \mathbf{H}^*(\mathbf{r})\right] \tag{S49}$$

The associated angular momentum is

$$\mathbf{J} = \sum_{n=1}^{6} \mathbf{r}_n \times \mathbf{S}(\mathbf{r}_n) = \mathrm{Re}\left\{\sum_{n=1}^{6} \mathbf{r}_n \times [\mathbf{E}(\mathbf{r}_n) \times \mathbf{H}^*(\mathbf{r}_n)]\right\} \tag{S50}$$

where $\{\mathbf{r}_m\}$ are positions of the cylinders. The $z$-component of $\mathbf{J}$ is

$$\begin{aligned}
J_z &= \sum_{n=1}^{6} r_n S_\varphi(\mathbf{r}_n) = -\mathrm{Re}\left\{\sum_{n=1}^{6} r_n E_\rho(\mathbf{r}_n) H_z^*(\mathbf{r}_n)\right\} \\
&= \mathrm{Re}\left\{\frac{1}{i\omega}\sum_{n=1}^{6}\left[H_z^*(\mathbf{r})\frac{\partial H_z(\mathbf{r})}{\partial \varphi}\right]_{\mathbf{r}=\mathbf{r}_n}\right\}
\end{aligned} \tag{S51}$$

where we used $r_n = \rho_n$.

In our experiment, the measured transmission at a position $\mathbf{r}$ is proportional to $H_z(\mathbf{r})$: $S_{21}(\mathbf{r}) \propto H_z(\mathbf{r})$. We thus have

$$J_z \propto \mathrm{Re}\left\{\frac{1}{i\omega}\sum_{n=1}^{6}\left[S_{21}^*(\mathbf{r})\frac{\partial S_{21}(\mathbf{r})}{\partial \varphi}\right]_{\mathbf{r}=\mathbf{r}_n}\right\} \propto \sum_{n=1}^{6} \mathrm{Im}\left[S_{21}^*(\mathbf{r}_n) S_{21}(\mathbf{r}_{n+1})\right] \tag{S52}$$

where we have approximated the derivative by a finite difference and $\mathbf{r}_7 = \mathbf{r}_1$.

## V. CALCULATION OF TOPOLOGICAL INVARIANTS

### A. Chern number

The Chern number [S6] is a topological invariant characterizing the topological properties of a band in a two-dimensional physical system. Consider a honeycomb lattice corresponding to $R = a$ in the main text. An eigenstate

of Hamiltonian $\hat{H}$ associated with the band $m$ is $\psi_{m,\mathbf{k}}(\mathbf{r}) = \varphi_{m,\mathbf{k}}(\mathbf{r})\exp(i\mathbf{k}\cdot\mathbf{r})$. Berry connection of the band $m$ is defined as

$$\mathbf{A}_m(\mathbf{k}) = i\langle \varphi_{m,\mathbf{k}} | \nabla_\mathbf{k} | \varphi_{m,\mathbf{k}} \rangle \tag{S53}$$

Berry curvature of the band $m$ is

$$\Omega_m(\mathbf{k}) = \partial_{k_x} A_{m,y}(\mathbf{k}) - \partial_{k_y} A_{m,x}(\mathbf{k}) = i\left(\left\langle \frac{\partial \varphi_{m,\mathbf{k}}}{\partial k_x} \middle| \frac{\partial \varphi_{m,\mathbf{k}}}{\partial k_y} \right\rangle - \left\langle \frac{\partial \varphi_{m,\mathbf{k}}}{\partial k_y} \middle| \frac{\partial \varphi_{m,\mathbf{k}}}{\partial k_x} \right\rangle\right) \tag{S54}$$

By integrating over the Brillouin zone, we obtain the Chern number of the band $m$:

$$C_m = \frac{1}{2\pi i} \int_{\text{BZ}} \Omega_m(\mathbf{k}) \mathrm{d}^2 \mathbf{k} \in \mathbb{Z} \tag{S55}$$

The Chern number of an energy gap is defined as a sum of Chern numbers of bands below the gap:

$$C = \sum_{m | E_m < E_{\text{gap}}} C_m \tag{S56}$$

where $E_{\text{gap}}$ is an energy inside the gap.

Equation (S55) can be used for computing the Chern number when analytical expressions of eigenstates are known. However, it is not convenient for numerical evaluation. We instead use a method proposed in Ref. [S7] that we briefly summarize below. Consider a rectangular discretization of the Brillouin zone illustrated in Fig. S8. For each $\mathbf{k}_j$, one introduces the following quantity:

$$U_\mu^m(\mathbf{k}_j) = \frac{\langle \varphi_{m,\mathbf{k}_j} | \varphi_{m,\mathbf{k}_j+\mathbf{u}_\mu} \rangle}{|\langle \varphi_{m,\mathbf{k}_j} | \varphi_{m,\mathbf{k}_j+\mathbf{u}_\mu} \rangle|}, \quad \mu = x, y \tag{S57}$$

where $\mathbf{u}_\mu$ are vectors connecting neighboring points of the discrete grid along axis $\mu = x$ or $\mu = y$, see Fig. S8. The lattice field strength is a discretized equivalent of the Berry curvature and is defined by

$$F^m(\mathbf{k}_j) = \ln\left[U_x^m(\mathbf{k}_j) U_y^m(\mathbf{k}_j) U_x^m(\mathbf{k}_j + \mathbf{u}_y)^{-1} U_y^m(\mathbf{k}_j + \mathbf{u}_x)^{-1}\right] \tag{S58}$$

The lattice field strength is summed over $\mathbf{k}_j$ in the discretized Brillouin zone to yield the Chern number of the band $m$:

$$C_m = \frac{1}{2\pi i} \sum_j F^m(\mathbf{k}_j) \tag{S59}$$

It can be demonstrated that Eq. (S59) is equivalent to Eq. (S55).

### B. Spin Chern number

Whereas the Chern number is a useful topological invariant for systems with broken time-reversal (TR) symmetry, it is always equal to zero for the system with preserved TR symmetry considered in the present work. Following the work done for characterizing topological properties of systems exhibiting the quantum spin Hall effect (QSHE) [S8–S10], we define a matrix

$$\hat{H}'(\mathbf{k}) = \hat{P}(\mathbf{k})\hat{J}_z\hat{P}(\mathbf{k}) \tag{S60}$$

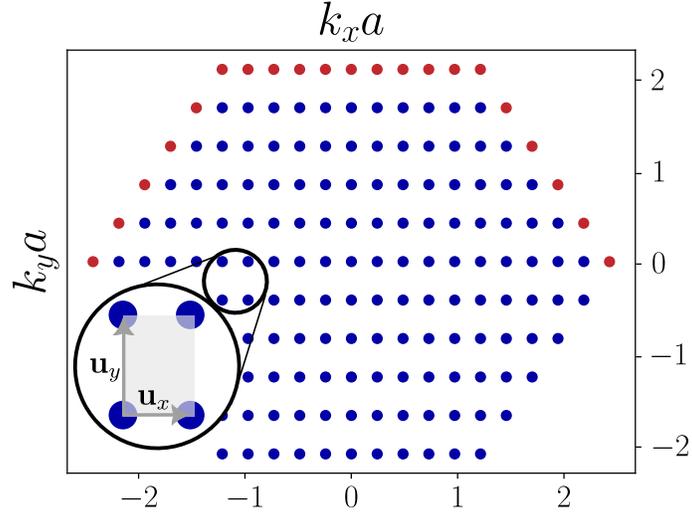

FIG. S8. Discretized Brillouin zone used to compute the Chern number. Red points are not included in the sum (S59) as they belong to other Brillouin zones. The ratio $u_x/u_y = 1/\sqrt{3}$ ensures that points of the discrete lattice follow the border of the Brillouin zone.

where

$$\hat{P} = \sum_{m|E_m < E_{\text{gap}}} |\varphi_{m,\mathbf{k}}\rangle\langle\varphi_{m,\mathbf{k}}| \tag{S61}$$

is the projector on the states with energies below the gap,

$$\hat{J}_z = \hat{U}\text{diag}(0, 1, -1, 1, -1, 0)\hat{U}^{-1} \tag{S62}$$

is a pseudospin operator, and

$$\hat{U} = (s, p_+, p_-, d_+, d_-, f) \tag{S63}$$

defines a transformation into "*spdf*" basis obtained by diagonalizing the Hamiltonian

$$\hat{H}_1 = \begin{pmatrix} 0 & 1 & 0 & 0 & 0 & 1 \\ 1 & 0 & 1 & 0 & 0 & 0 \\ 0 & 1 & 0 & 1 & 0 & 0 \\ 0 & 0 & 1 & 0 & 1 & 0 \\ 0 & 0 & 0 & 1 & 0 & 1 \\ 1 & 0 & 0 & 0 & 1 & 0 \end{pmatrix} \tag{S64}$$

of a hexagonal six-site cluster (without loss of generality, we set coupling between nearest-neighbor sites $t_{\text{in}} = 1$). The spectrum of $\hat{H}_1$ is

$$\text{Spectrum}(\hat{H}_1) = \{-2, -1, -1, 1, 1, 2\} \tag{S65}$$

and the corresponding eigenvectors are

$$s = \frac{1}{\sqrt{6}}(1,1,1,1,1,1)^T$$
$$p_x = \frac{1}{2}(1,1,0,-1,-1,0)^T$$
$$p_y = \frac{1}{\sqrt{12}}(1,-1,-2,-1,1,2)^T$$
$$d_{x^2-y^2} = \frac{1}{\sqrt{12}}(1,1,-2,1,1,-2)^T$$
$$d_{xy} = \frac{1}{2}(1,-1,0,1,-1,0)^T$$
$$f = \frac{1}{\sqrt{6}}(1,-1,1,-1,1,-1)^T \tag{S66}$$

These eigenstates resemble standard atomic orbitals [S11]. To obtain basis states carrying well-defined pseudospins, we transform $p_x$, $p_y$, $d_{x^2-y^2}$ and $d_{xy}$ orbitals into

$$p_\pm = \frac{1}{\sqrt{2}}(p_x \pm i p_y), \quad d_\pm = \frac{1}{\sqrt{2}}(d_{x^2-y^2} \pm i d_{xy}) \tag{S67}$$

The use of *spdf* basis to characterize topological phenomena in systems with preserved TR symmetry has been first proposed by Wu and Hu [S12, S13]. One can verify that vectors $s$, $p_+$, $p_-$, $d_+$, $d_-$, $f$ are eigenvectors of the pseudospin operator $\hat{J}_z$ with eigenvalues 0, 1, −1, 1, −1, 0 given by the diagonal matrix in Eq. (S62).

With the definitions above, and if the different six-site clusters composing an infinite lattice are not coupled, $\hat{H}'(\mathbf{k})$ has the eigenvectors $|\varphi_{m,\mathbf{k}}\rangle$ of $\hat{H}$ corresponding to eigenenergies $E_m < E_{\text{gap}}$ but eigenvalues of $\hat{J}_z$. In the presence of coupling between different six-site clusters that is necessary to model our experimental situation, eigenvalues of $\hat{H}'(\mathbf{k})$ change but keep their signs unchanged as far as coupling is not very strong. This property allows for separating eigenstates that may be degenerate in energy but have different signs of pseudospin. Chern numbers can now be computed separately for states with positive and negative pseudospins by the algorithm described in Sec. V A (more precisely, Eqs. (S57), (S58) and (S59)) using $\hat{H}'$ instead of $\hat{H}$.[S14] The difference of Chern numbers $C_\pm$ of "bands" with positive and negative eigenvalues of $\hat{H}'$ yields the spin Chern number

$$C_{\text{SC}} = \frac{1}{2}(C_+ - C_-) \tag{S68}$$

The spin Chern number for the system considered in the present work is shown in Fig. 1(d) of the main text. Note that $C_+ + C_- = C = 0$.

### C. Bott index

The Bott index has been introduced for characterization of physical systems by Loring and Hastings [S15, S16]. It is useful to characterize topological properties of systems of finite size as well as in the presence of disorder, for which calculation of Chern number is not straightforward [S10, S17].

Consider a tight-binding model on a rectangular lattice of size $L_x \times L_y$ with periodic boundary conditions (PBC). Lattice sites are

$$\mathbf{r}_i = \{x_i, y_i\}, \quad i = 1, \ldots, N \tag{S69}$$

The eigenbasis $\{\boldsymbol{\psi}_1, \ldots, \boldsymbol{\psi}_N\}$ of $\hat{H}$ is split in two parts: $\{\boldsymbol{\psi}_1, \ldots, \boldsymbol{\psi}_m\}$ generates the subspace below a certain energy $E$ whereas $\{\boldsymbol{\psi}_{m+1}, \ldots, \boldsymbol{\psi}_N\}$ generates the subspace above $E$. We define an $m \times N$ matrix

$$\hat{W} = \{\boldsymbol{\psi}_1, \ldots, \boldsymbol{\psi}_m\} \tag{S70}$$

diagonal position matrices

$$\hat{X} = \mathrm{diag}(x_1, \ldots, x_N), \quad \hat{Y} = \mathrm{diag}(y_1, \ldots, y_N) \tag{S71}$$

and

$$\hat{U} = \hat{W}^\dagger \exp\left(i\frac{2\pi}{L_x}\hat{X}\right) \hat{W}, \quad \hat{V} = \hat{W}^\dagger \exp\left(i\frac{2\pi}{L_y}\hat{Y}\right) \hat{W} \tag{S72}$$

The Bott index is

$$C_{\mathrm{B}}(E) = \frac{1}{2\pi} \mathrm{Im}\,\mathrm{Tr}\left[\ln\left(\hat{V}\hat{U}\hat{V}^{-1}\hat{U}^{-1}\right)\right] \tag{S73}$$

If $\hat{U}$ and $\hat{V}$ are almost unitary: $UU^\dagger \approx VV^\dagger \approx \mathbb{1}$, one can use another definition:

$$C_{\mathrm{B}}(E) = \frac{1}{2\pi} \mathrm{Im}\,\mathrm{Tr}\left[\ln\left(\hat{V}\hat{U}\hat{V}^\dagger\hat{U}^\dagger\right)\right] \tag{S74}$$

which is easier to implement numerically.

A way to evaluate Eqs. (S73) and (S74) is to find eigenvalues $\Lambda_j$ of the matrix under the logarithm and then sum their logarithms:

$$C_{\mathrm{B}}(E) = \frac{1}{2\pi} \mathrm{Im} \sum_j \ln(\Lambda_j) \tag{S75}$$

Remarkably, the result is always an integer and it converges to the Chern number for $E$ inside a band gap of infinite lattice to which the considered finite lattice tends in the limit of $L_x, L_y \to \infty$ [S18].

An alternative but equivalent definition of Bott index uses the projector

$$\hat{P} = \sum_{j \leqslant m} |\boldsymbol{\psi}_j\rangle\langle\boldsymbol{\psi}_j| \tag{S76}$$

onto eigenstates below $E$ instead of $W$. With definitions

$$\hat{U} = \hat{P}\mathrm{e}^{2i\pi\hat{X}/L_x}\hat{P}, \quad \hat{V} = \hat{P}\mathrm{e}^{2i\pi\hat{Y}/L_x}\hat{P} \tag{S77}$$

$C_{\mathrm{B}}$ is again given by Eqs. (S73) or (S74).

### D. Spin Bott index

The spin Bott index has been introduced in Refs. [S19, S20] by analogy with the spin Chern number to characterize topological properties of systems with preserved TR symmetry and, in particular, those exhibiting QSHE. We adapt its definition to our situation of tight-binding model in which $N$ sites of a two-dimensional lattice are arranged in $N/6$ hexagonal six-site clusters with inter-cluster coupling $t_{\mathrm{in}}$ and inter-cluster coupling $t_{\mathrm{out}}$. Following the same steps as in Sec. V B but now in real space, we define a projector on states below en energy $E_{\mathrm{gap}}$ inside the energy gap

$$\hat{P} = \sum_{m|E_m < E_{\mathrm{gap}}} |\boldsymbol{\psi}_m\rangle\langle\boldsymbol{\psi}_m| \tag{S78}$$

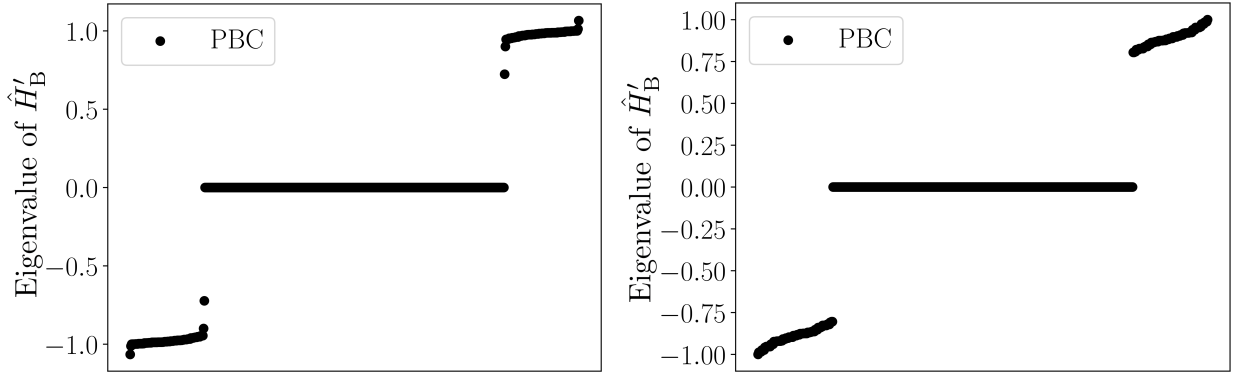

FIG. S9. Spectrum of $\hat{H}'_\text{B}$ for $t_\text{in} = 1$ and $t_\text{out} = 0.5$ (left) or 1.5 (right), for a lattice with periodic boundary conditions (PBC) and $N = 216$.

an $N \times N$ block-diagonal matrix

$$\hat{U}_\text{B} = \begin{pmatrix} \hat{U} & & \\ & \ddots & \\ & & \hat{U} \end{pmatrix} \tag{S79}$$

where the $6 \times 6$ matrix $\hat{U}$ is given by Eq. (S63), a pseudospin operator

$$\hat{J}_z^{(\text{B})} = U_\text{B} \text{diag}(0, 1, -1, 1, -1, 0, \ldots, 0, 1, -1, 1, -1, 0) U_\text{B}^{-1} \tag{S80}$$

where the diagonal matrix in the middle is obtained by repeating the same $6 \times 6$ diagonal matrix $\text{diag}(0, 1, -1, 1, -1, 0)$ along the diagonal $N/6$ times, and

$$\hat{H}'_\text{B} = \hat{P} \hat{J}_z^{(\text{B})} \hat{P} \tag{S81}$$

$\hat{H}'_\text{B}$ is constructed in such a way that in the absence of coupling between six-atom clusters, its eigenvalues are equal to those of $\hat{J}_z^{(\text{B})}$ and are either $\pm 1$ or $0$. The corresponding eigenvectors are $p_+$, $d_+$ (eigenvalues $J_z^{(\text{B})} = +1$), $p_-$, $d_-$ (eigenvalues $J_z^{(\text{B})} = -1$), $s$ or $f$ (eigenvalues $J_z^{(\text{B})} = 0$) repeated on each six-atoms cluster. Using $\hat{H}'_\text{B}$ instead of $\hat{H}$, allows for calculating the Bott indices $C_\text{B}^\pm$ of each of "bands" corresponding to $J_z^{(\text{B})} = \pm 1$ by following the procedure described in Sec. V C. If different six-atom clusters are coupled, the eigenvalues of $\hat{H}'_\text{B}$ are not equal to $\pm 1$ and $0$ anymore (see Fig. S9) and the eigenvectors differ from $s$, $p$, $d$ and $f$ states. However, as long as the couppling is not too strong, we can still compute Bott indices $C_\text{B}^\pm$ by taking into account states with positive ($C_\text{B}^+$) or negative ($C_\text{B}^-$) eigenvalues $J_z^{(\text{B})}$ and ignoring the states with $J_z^{(\text{B})} = 0$. Their difference yields the spin Bott index

$$C_\text{SB} = \frac{1}{2} \left( C_\text{B}^+ - C_\text{B}^- \right) \tag{S82}$$

that is shown in Fig. 1(d) of the main text for different sample sizes and different types of boundary conditions. Note that in our system, $C_\text{B}^+ + C_\text{B}^- = C_\text{B} = 0$.

The calculation of spin Bott index is performed using PyBott software package [S21] .

---


\fi

\end{document}